\documentclass{llncs}
\usepackage{txfonts}
\usepackage{algorithm}
\usepackage{algorithmic}
\usepackage{graphicx}

\newtheorem{Def}{Definition}
\newtheorem{Exm}{Example}

\title{Making Abstraction Refinement Efficient in Model Checking\thanks{This research is supported by
the NSFC Grant No. 60373103, 60433010, 60873018 and 60910004, DPRPC
Grant No. 51315050105, 973 Program Grant No. 2010CB328102 and SRFDP
Grant No. 200807010012.}}

\author{Cong Tian and Zhenhua Duan}

\institute{Institute of Computing Theory \& Technology, Xidian
University, Xi'an, 710071, P.R. China\\
\{ctian, zhhduan\}@mail.xidian.edu.cn}

\begin{document}
\maketitle
\begin{abstract}

 Abstraction is one of the most important strategies for dealing with the state space explosion problem in model
 checking. In the abstract model, although the state space is largely reduced, however, a counterexample found in
 such a model may not be a real counterexample. And the abstract model needs to be further refined where
 an NP-hard state separation problem is often involved.
In this paper, a novel method is presented by adding extra variables
to the abstract model for the refinement. With this method, not only
the NP-hard state separation problem is avoided, but also a smaller
refined abstract model is obtained.
\end{abstract}

\section{Introduction}
Model checking is an important approach for the verification of
hardware, software, multi-agent systems, communication protocols,
embedded systems and so forth. The term model checking was coined by
Clarke and Emerson \cite{Clarke81}, as well as Sifakis and Queille
\cite{QS82} independently. The earlier model checking algorithms
explicitly enumerated the reachable states of the system in order to
check the correctness of a given specification. This restricted the
capacity of model checkers to systems with a few million states.
Since the number of states can grow exponentially in the number of
variables, early implementations were only able to handle
 small designs and did not scale to examples with industrial complexity. To combat this, kinds of methods, such as abstraction,
 partial order reduction, OBDD, symmetry and bound technique are applied to model checking to reduce the state space for efficient
 verification. Thanks to these efforts, model checking has been one of the most successful verification approaches which is widely
 adopted in the industrial community.

Among the techniques for reducing the state space, abstraction is
certainly the most important one. Abstraction technique preserves
all the behaviors of the concrete system but may introduce behaviors
that are not present originally. Thus, if a property (i.e. a
temporal logic formula) is satisfied in the abstract model, it will
still be satisfied in the concrete model. However, if a property is
unsatisfiable in the abstract model, it may still be satisfied in
the concrete model, and none of the behaviors that violate the
property in the abstract model can be reproduced in the concrete
model. In this case, the counterexample is said to be spurious.
Thus, when a spurious counterexample is found, the abstraction
should be refined in order to eliminate the spurious behaviors. This
process is repeated until either a real counterexample is found or
the abstract model satisfies the property.

There are many techniques for generating the initial abstraction and
refining the abstract models. We follow the counterexample guided
abstraction and refinement method proposed by Clarke, etc
\cite{Clarke04}. With this method, abstraction is performed by
selecting a set of variables which are insensitive to the desired
property to be invisible. In each iteration, a model checker is
employed to check whether or not the abstract model satisfies the
desired property. If a counterexample is reported, it is simulated
with the concrete model by a SAT solver or checked by other
algorithms. Then, if the counterexample is checked to be spurious, a
set of invisible variables are made visible to refine the abstract
model. With this method, to find the coarsest (or smallest) refined
model is NP-hard \cite{CGJLV00}. Further, it is important to find a
small set of variables in order to keep the size of the abstract
state space smaller. However, to find the smallest set of variables
is also NP-hard \cite{HSNH10}. To combat this, Integer Linear
Program (ILP) based separation algorithm which outputs the minimal
separating set is given \cite{Clarke04}. And a polynomial
approximation algorithm based on Decision Trees Learning (DTL) is
also presented \cite{Clarke04}. Moreover, Heuristic-Guided
separating algorithms are presented in \cite{HSGS09}, and
evolutional algorithms are introduced in \cite{HSNH10} for the state
separation problem. These approximate algorithms are compared with
experimental results.

In this paper, we follow the abstract method used in
\cite{Clarke04,HSGS09,HSNH10} by selecting some set of variables to
be invisible. Then we evaluate the counterexample with Algorithm
{\scshape CheckSpurious}. When a failure state is achieved, instead
of selecting some invisible variables to be visible, extra variables
are added to the abstract model for the refinement. With this
method, not only the NP-hard state separation problem is avoided,
but also a smaller refined abstract model is obtained.

The rest parts of the paper are organized as follows. The next
section briefly presents the related work
 concerning abstraction
refinement in model checking. In section 3, the abstraction
algorithm is formalized by making insensitive variables invisible.
In section 4, by formally defining spurious counterexamples, the
algorithm for checking whether or not a counterexample in the
abstract model is spurious is presented. Further, the new
abstraction refinement algorithm is given. Subsequently, abstraction
model checking framework based on the new proposed algorithms is
illustrated in section 5. Finally, conclusions are drawn in section
6.
\section{Related Work}
We focus on the Counter-Example Guided Abstraction Refinement
(CEGAR) framework which was fist proposed by Kurshan
\cite{Krushan94}. Recently, some variations of the basic CEGAR were
given \cite{Clarke04,WLJHS06,CGHS02,CCKSVW02,HJMS02,GKMFV03,GD00}.
Most of them use a model checker and try to get rid of spurious
counterexamples to achieve a concrete counterexample or a proof of
the desired property.

The closest works to ours are those where the abstract models are
obtained by making some of the variables invisible. To the best of
our knowledge, this abstraction method was first proposed by Clarke,
etc. \cite{Clarke04,CGHS02}. With their approach, abstraction is
performed by selecting a set of variables (or latches in circuits)
to be invisible. In each iteration, a standard Ordered Binary
Decision Diagram (OBDD)-based symbolic model checker is used to
check whether or not the abstract model satisfies the desired
property which is described by a formula in temporal logic. If a
counterexample is reported by the model checker, it is simulated
with the concrete system by a SAT solver. It tells us that the model
is satisfiable if the counterexample is a real one, otherwise, the
counterexample is a spurious one and a failure state is found which
is the the last state in the longest prefix of the counterexample
that is still satisfiable. Subsequently, the failure state is used
to refine the abstraction by making some invisible variables
visible. With this method, to find the smallest refined model is
NP-hard \cite{CGJLV00}. To combat this, both optimal exponential and
approximate polynomial algorithms are given. The first one is done
by using an ILP solver which is known to be NP complete; and the
second one is based on machine learning approaches.

Some heuristics for refinement variables selection were first
presented in \cite{HSGS09}. It studied on effective greedy heuristic
algorithms on state separation problem. Further, in \cite{HSHGS10},
probabilistic learning approach which utilized the sample learning
technique, evolutionary algorithm and effective heuristics were
proposed. The performances were illustrated by experiment results.

\section{Abstraction Function}
As usual, a Kripke structure \cite{Krip63} is used to model a
system. Let $V=\{v_1,...,v_n\}$ ranging over a finite domain
$D\cup\{\bot\}$ be the set of variables involved in a system. For
any $v_i\in V$, $1\leq i\leq n$, a set of the valuations of $v_i$ is
defined by,
$$\Sigma_{v_i}= \{v_i=d\mid d\in D\cup\{\bot\}\}$$
where $v_i=\bot$ means $v_i$ is undefined. Further, the set of all
the possible states of the system, $\Sigma$,  is defined by,
$$\Sigma= \Sigma_{v_1}\times ...\times \Sigma_{v_n}$$
Let $\mathit{AP}$ be the set of propositions. A Kripke structure
over $AP$ is a tuple $K=(S,S_0,R,L)$, where $S\subseteq \Sigma$ is
the set of states (i.e. a state in $S$ is a valuation of variables
in $V$), $S_0\subseteq S$ is the set of initial states, $R\subseteq
S\times S$ is the transition relation, $L:S\rightarrow 2^{AP}$ is
the labeling function. For convenience, $s(v)$ is employed to denote
the value of $v$ at state $s$. A path in a Kripke structure is a
sequence of states, $\Pi=s_1, s_2, ...$, where $s_1\in S_0$ and
$(s_i,s_{i+1})\in R$ for any $i\geq 1$.

Following the idea given in \cite{Clarke04}, we separate $V$ into
two parts $V_V$ and $V_I$ with $V=V_V\cup V_I$. $V_V$ stands for the
set of visible variables while $V_I$ denotes the set of invisible
variables. Invisible variables are those that we do not care about
and will be ignored when building the abstract model. In the
original model $K=(S,S_0,R,L)$, all variables are visible ($V_V=V$,
$V_I=\emptyset$). To obtain the abstract model
$\hat{K}=(\hat{S},\hat{S_0},\hat{R},\hat{L})$, some variables, e.g.
$V_X\subseteq V$, are selected to be invisible ($V_V=V\setminus
V_X$, $V_I=V_X$). Thus, the set of all possible states in the
abstract model will be:
$$\hat{\Sigma}=\Sigma_{v_1}\times ...\times\Sigma_{v_k}$$ where $k=|V_V|< n$,
and for each $1\leq i\leq k$, $v_i\in V_V$. That is
$\hat{S}\subseteq \hat{\Sigma}$. For a state $s\in S$ and a state
$\hat{s}\in \hat{S}$, we say $\hat{s}$ is the projection of $s$ in
the abstract model by making $V_V$ visible, denoted by $h(s,V_V)$,
iff $s(v)=\hat{s}(v)$ for any $v\in V_V$. Inversely, $s$ is called
the origin of $\hat{s}$, and the set of origins of $\hat{s}$ is
denoted by $h^-(\hat{s}, V_V)$.

Therefore, given the original model $K=(S,S_0,R,L)$ and the the
selected visible variables $V_V$, the abstract model
$\hat{K}=(\hat{S},\hat{S_0},\hat{R},\hat{L})$ can be obtained by
Algorithm {\scshape Abstract} as shown below.

\begin{algorithm}[h]
\caption{: {\scshape Abstract}$(K,V_V)$}
  {\bf Input}: the original model $K=(S,S_0,R,L)$ and a set of selected visible variables $V_V$\\
  {\bf Output}: the abstract model
  $\hat{K}$$=$$(\hat{S}$,$\hat{S_0}$,$\hat{R}$,$\hat{L})$

    \begin{algorithmic}[1]
\STATE  $\hat{S}=\{\hat{s}\in \hat{\Sigma}~|\mbox{ there exists
}s\in S\mbox{ such that }h(s,V_V)=\hat{s}\}$;

\STATE $\hat{S_0}=\{\hat{s}\in \hat{S}~|\mbox{ there exists }s\in
S_0\mbox{ such that }h(s,V_V)= \hat{s}\}$;

\STATE $\hat{R}=\{(\hat{s_1},\hat{s_2})\mid \hat{s_1},\hat{s_2}\in
\hat{S}, \mbox{and there exist }s_1,s_2\in S\mbox{ such that
}h(s_1,V_V)=\hat{s_1}, h(s_2,V_V) = \hat{s_2} \mbox{ and
}$\\$(s_1,s_2)\in R\}$;

\STATE $L(\hat{s})=\bigcup\limits_{s\in S, h(s,V_V)=\hat{s}} L(s)$;

\STATE return $\hat{K}=(\hat{S},\hat{S_0},\hat{R},\hat{L})$;
    \end{algorithmic}
\end{algorithm}

\begin{Exm}\rm As illustrated in Figure\ref{fig:Af}, the original
model is a Kripke structure with four states.
\begin{figure}[htp]
\centerline{\includegraphics[height=4.3cm]{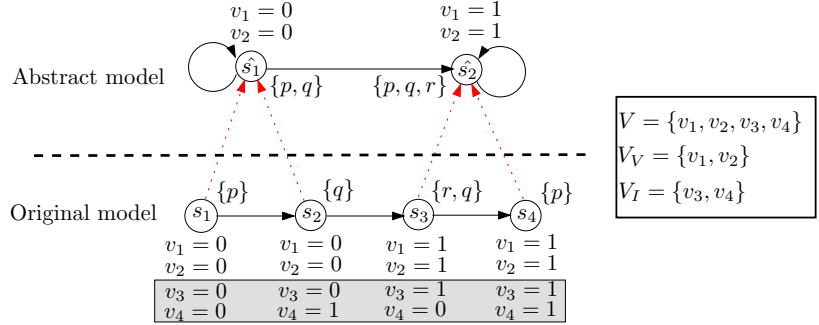}}\caption{Abstraction}\label{fig:Af}
\end{figure}
Initially, the system has four variables $v_1$, $v_2$, $v_3$ and
$v_4$. Assume that  $v_3$ and $v_4$ are selected to be invisible. By
Algorithm {\scshape Abstract}, an abstract model with two states is
obtained. In the abstract model, $\hat{s_1}$ is the projection of
$s_1$ and $s_2$, while $\hat{s_2}$ is the projection of $s_3$ and
$s_4$. $(\hat{s_1},\hat{s_2})\in \hat{R}$ since $(s_2,s_3)\in R$,
and $(\hat{s_1},\hat{s_1})$, $(\hat{s_2},\hat{s_2})\in \hat{R}$
because of $(s_1,s_2)$, $(s_3,s_4)\in R$.\hfill{$\Box$}
\end{Exm}

\section{Refinement}
\subsection{Why Refining?}
It can be observed that the state space is largely reduced in the
abstract model. However, when implementing model checking with the
abstract model, some reported counterexamples will not be real
counterexamples that violate the desired property, since the
abstract model contains more paths than the original model. This is
further illustrated in the traffic lights controller example given
below. The example was first presented in \cite{CGJLV00}.
\begin{Exm}\rm
For the traffic light controller in Figure \ref{fig:light}, we want
to prove $\Box\Diamond (state=stop)$ (any time, the state of the
light will be $stop$ sometimes in the future). By implementing model
checking with the abstract model in the right hand side of Figure
\ref{fig:light} where the variable $color$ is made invisible,
\begin{figure}[htp]
\centerline{\includegraphics[height=3.5cm]{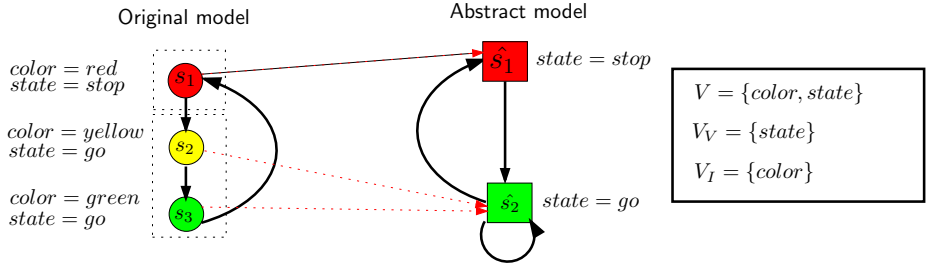}}\caption{Traffic
Light Controller}\label{fig:light}
\end{figure}
 a counterexample, $\hat{s_1},\hat{s_2},\hat{s_2},\hat{s_2},...$ will be reported. However, in the concrete model, such a behavior cannot be found. So, this is not a real counterexample. \hfill{$\Box$}
\end{Exm}

\subsection{Spurious Counterexamples}
\begin{figure}[htp]
\centerline{\includegraphics[height=6cm]{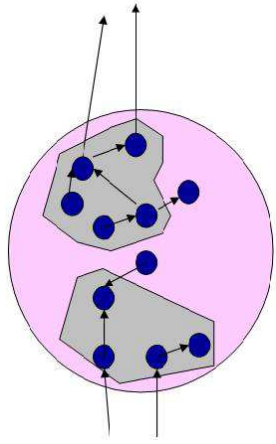}}\caption{$In_{\hat{s_i}}$
and $Out_{\hat{s_i}}$}\label{fig:inout}
\end{figure}

As pointed in \cite{Clarke04,HSHGS10}, a counterexample in the
abstract model which does not exist in the concrete model is called
a spurious counterexample. To formally define a spurious
counterexample, we first introduce failure states.
To this end, $In_{\hat{s_i}}^0$, $In_{\hat{s_i}}^1$, ...,
$In_{\hat{s_i}}^n$ and  $In_{\hat{s_i}}$ are defined first:
$$
\begin{array}{llll}
In_{\hat{s_i}}^0&=&\{s\mid s\in h^-(\hat{s_i},V_V), s'\in h^-(\hat{s_{i-1}},V_V) \mbox{ and }\\&&(s',s)\in R \}\\
In_{\hat{s_i}}^1&=&\{s\mid s\in h^-(\hat{s_i},V_V), s'\in In_{\hat{s_i}}^0 \mbox{ and }(s',s)\in R \}\\
&...&\\
In_{\hat{s_i}}^n&=&\{s\mid s\in h^-(\hat{s_i},V_V), s'\in In_{\hat{s_i}}^{n-1} \mbox{ and }(s',s)\in R \}\\
&...&\\
In_{\hat{s_i}}&=&\bigcup\limits_{i=0}^\infty In_{\hat{s_i}}^i\\
\end{array}
$$
Clearly, $In_{\hat{s_i}}^0$ denotes the set of states in
$h^-(\hat{s_i},V_V)$ with inputting edges from the states in
$h^-(\hat{s_{i-1}},V_V)$, and $In_{\hat{s_i}}^1$ stands for the set
of states in $h^-(\hat{s_i},V_V)$ with inputting edges from the
states in $In_{\hat{s_i}}^0$, and $In_{\hat{s_i}}^2$ means the set
of states in $h^-(\hat{s_i},V_V)$ with inputting edges from the
states in $In_{\hat{s_i}}^1$, and so on. Thus, $In_{\hat{s_i}}$
denotes the set of states in $h^-(\hat{s_i},V_V)$  that are
reachable from some state in $h^-(\hat{s_{i-1}},V_V)$ as illustrated
in the lower gray part in Figure \ref{fig:inout}. Note that there
must exist a natural number $n$, such that
$\bigcup\limits_{i=0}^{n+1}In_{\hat{s_i}}^{i}
=\bigcup\limits_{i=0}^{n}In_{\hat{s_i}}^{i}$ since
$h^-(\hat{s_i},V_V)$ is finite.
Similarly,  $Out_{\hat{s_i}}^0$, $Out_{\hat{s_i}}^1$, ...,
$Out_{\hat{s_i}}^n$  and $Out_{\hat{s_i}}$ can also be defined.
$$
\begin{array}{llll}
Out_{\hat{s_i}}^0&=&\{s\mid s\in h^-(\hat{s_i},V_V), s'\in h^-(\hat{s_{i+1}},V_V) \mbox{ and }\\&&(s,s')\in R \}\\
Out_{\hat{s_i}}^1&=&\{s\mid s\in h^-(\hat{s_i},V_V), s'\in Out_{\hat{s_i}}^0 \mbox{ and }(s,s')\in R \}\\
&...&\\
Out_{\hat{s_i}}^n&=&\{s\mid s\in h^-(\hat{s_i},V_V), s'\in Out_{\hat{s_i}}^{n-1} \mbox{ and }(s,s')\in R \}\\
&...&\\
Out_{\hat{s_i}}&=&\bigcup\limits_{i=0}^\infty Out_{\hat{s_i}}^i\\
\end{array}
$$
Where $Out_{\hat{s_i}}^0$ denotes the set of states in
$h^-(\hat{s_i},V_V)$ with outputting edges to the states in
$h^-(\hat{s_{i+1}},V_V)$, and $Out_{\hat{s_i}}^1$ stands for the set
of states in $h^-(\hat{s_i},V_V)$ with outputting edges to the
states in $Out_{\hat{s_i}}^0$, and $Out_{\hat{s_i}}^2$ means the set
of states in $h^-(\hat{s_i},V_V)$ with outputting edges to the
states in $Out_{\hat{s_i}}^1$, and so on. Thus, $Out_{\hat{s_i}}$
denotes the set of states in $h^-(\hat{s_i},V_V)$ from which some
state in $h^-(\hat{s_{i+1}},V_V)$ are reachable as depicted in the
higher gray part in Figure \ref{fig:inout}. Similar to
$In_{\hat{s_i}}$, there must exist a natural number $n$, such that
$\bigcup\limits_{i=0}^{n+1}Out_{\hat{s_i}}^{i}
=\bigcup\limits_{i=0}^{n}Out_{\hat{s_i}}^{i}$. Accordingly, a
failure state can be defined as follows.

\begin{Def}{\bf (Failure States) }\rm A state $\hat{s_i}$ in a counterexample $\hat{\Pi}$ is a failure state if $In_{\hat{s_i}}\not=\emptyset$, $Out_{\hat{s_i}}\not=\emptyset$ and  $In_{\hat{s_i}}\cap Out_{\hat{s_i}}=\emptyset$. \hfill{$\Box$}
\end{Def}
Further, given a failure state $\hat{s_i}$ in a counterexample
$\hat\Pi$, the set of the origins of $\hat{s_i}$,
$h^-(\hat{s_i},V_V)$, is separated into three sets,
$\mathcal{D}=In_{\hat{s_i}}$ (the set of dead states),
$\mathcal{B}=Out_{\hat{s_i}}$ (the set of bad states) and
$\mathcal{I}=h^-(\hat{s_i})\setminus(\mathcal{D}\cup \mathcal{B})$
(the set of the isolated states). Note that by the definition of
failure state, $\mathcal{D}$ and $\mathcal{B}$ cannot be empty sets,
while $\mathcal{I}$ may be empty.

\begin{Def}{\bf (Spurious Counterexamples) }\rm A counterexample $\hat{\Pi}$ in an abstract model $\hat{K}$ is spurious
if there exists at least one failure state $\hat{s_i}$ in
$\hat{\Pi}$ \hfill{$\Box$}
\end{Def}

\begin{Exm}\rm
Figure \ref{fig:spurious} shows a spurious counterexample where the
state $\hat{3}$ is a failure  state.
\begin{figure}[htp]
\centerline{\includegraphics[height=3.5cm]{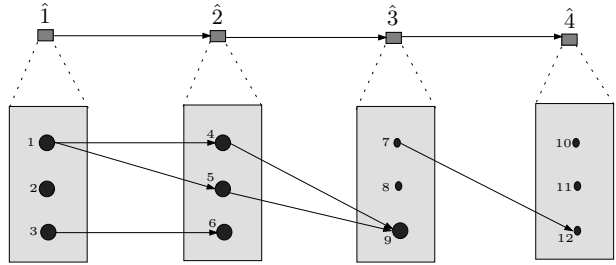}}\caption{A
Spurious Path}\label{fig:spurious}
\end{figure}
In the set, $h^-(\hat{3},V_V)=\{7,8,9\}$, of the origins of state
$\hat{3}$, $9$ is  a deadend state, $7$ is a bad state, and $8$ is
an isolated state. \hfill{$\Box$}
\end{Exm}

In \cite{CGJLV00}, Algorithm {\scshape SplitPath} is presented for
checking whether or not a counterexample is spurious. And in
\cite{Clarke04}, a SAT solver is used to check the counterexample.
We also present Algorithm {\scshape CheckSpurious} for checking
whether or not a counterexample is spurious based on the formal
definition of spurious paths. The algorithm takes a counterexample
as input and outputs the first failure state as well as
$\mathcal{D}$, $\mathcal{B}$ and  $\mathcal{I}$ with respect to the
failure state. Note that a counterexample may be a finite path
$<s_1,s_2,...,s_n>$, $n\geq 1$, or an infinite path
$<s_1,s_2,...,(s_i,...,s_j)^\omega>$, $1\leq i\leq j$, with a loop
suffix (a suffix produced by a loop). For the finite counterexample,
it will be checked directly while for an infinite one, we need only
check its finite prefix such as $<s_1,s_2,...,s_i,...,s_j,s_i>$.
\begin{algorithm}[h]
\caption{: {\scshape CheckSpurious}($\hat{\Pi}$)}
  {\bf Input}: a counterexample $\hat{\Pi}=<\hat{s_1},\hat{s_2},...,\hat{s_n}>$ in the abstract model $\hat{K}=(\hat{S},\hat{S_0},\hat{R},\hat{L})$, and the original model $K=(S,S_0,R,L)$\\
  {\bf Output}: a failure state $s_f$, $\mathcal{D}$, $\mathcal{B}$ and $\mathcal{I}$
    \begin{algorithmic}[1]
\STATE {\bf Initialization}: $int$ $i=2$;
        \WHILE {$i\leq n-1$}
            \STATE  {\bf if} $In_{\hat{s_i}}\cap Out_{\hat{s_i}}\not=\emptyset$, $i=i+1$;
            \STATE  {\bf else} return $s_f=\hat{s_{i}}$, $\mathcal{D}=In_{\hat{s_i}}$, $\mathcal{B}=Out_{\hat{s_i}}$, and  $\mathcal{I}=h^-(\hat{s_{i}})\setminus (\mathcal{B}\cup\mathcal{D})$; break;
        \ENDWHILE \label{code:recentEnd}
\STATE {\bf if} $i=n$, return $\hat{\Pi}$ is a real counterexample;
    \end{algorithmic}
\end{algorithm}

Compared with Algorithm {\scshape SplitPath}, to check whether or
not a state $\hat{s_i}$ is a failure state, it only relies on its
pre and post states, $\hat{s_{i-1}}$ and $\hat{s_{i+1}}$; while in
Algorithm {\scshape CheckSpurious}, to check state $\hat{s_i}$, it
relies on all states in the prefix, $\hat{s_1},...,\hat{s_{i-1}}$,
of $\hat{s_i}$. Based on this, to check a periodic infinite
counterexample, several repetitions of the periodic parts are
needed. In contrast, this can be easily done by checking the finite
prefix $<s_1,s_2,...,s_i,...,s_j,s_i>$ by Algorithm {\scshape
CheckSpurious}.
\subsection{Refining Algorithm}
When a failure state and the corresponding $\mathcal{D}$,
$\mathcal{B}$ and $\mathcal{I}$ are reported by Algorithm {\scshape
CheckSpurious}, we need further refine the abstract model such that
$\mathcal{D}$ and $\mathcal{B}$ are separated into different
abstract states. This can be achieved by making a set of invisible
variables, $U\subseteq V_I$, visible \cite{Clarke04}. With this
method, to find the coarsest refined model is NP-hard. Further, to
keep the size of the refined abstract state space smaller, it is
important to make $U$ as small as possible. However, to find the
smallest $U$ is also NP-hard \cite{HSHGS10}. In \cite{Clarke04}, an
ILP solver is used to obtain the minimal set. However, it is
inefficient when the problem size is large, since IPL is an NPC
problem. To combat this, several approximate polynomial algorithms
are proposed \cite{Clarke04,HSGS09,HSNH10} with non-optimal results.
Moreover, even though a coarser refined abstract model may be
produced by making $U$ smaller, it is uncertain that the smallest
$U$ will induce the coarsest refined abstract model. Motivated by
this, a new refinement approach is proposed by adding extra boolean
variables to the set of visible variables. With this approach,  not
only the NP-hard problem can be avoided but also a coarser refined
abstract model can be obtained. The basic idea for the refining
algorithm is described below.

\begin{figure}[htp]
\centerline{\includegraphics[height=4.5cm]{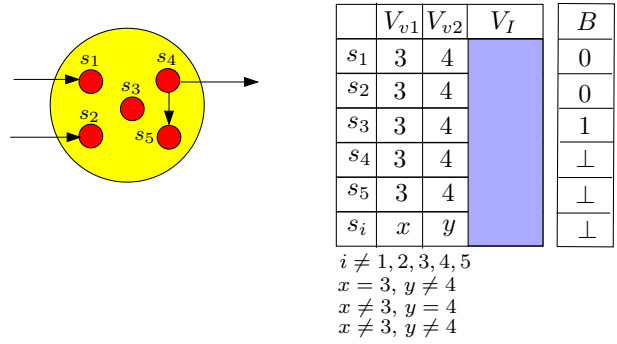}}\caption{A
Failure State}\label{fig:failure}
\end{figure}

\begin{figure}[htp]
\centerline{\includegraphics[height=5.3cm]{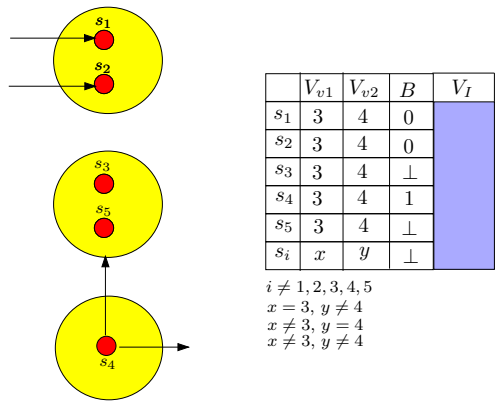}}\caption{Refined
Abstract States}\label{fig:refine}
\end{figure}

Assume that a failure state is found with $\mathcal{D}=\{s_1,s_2\}$,
$\mathcal{B}=\{s_4\}$ and $\mathcal{I}=\{s_3,s_5\}$ as illustrated
in Figure \ref{fig:failure} where the abstract model is obtained by
making $V_{v1}$ and $V_{v2}$ visible and other variables invisible.
To make $\mathcal{D}$ and $\mathcal{B}$ separated into two abstract
states,
 an extra  boolean variable $B$ is added to the system with the valuation being $0$ at the states in $\mathcal{D}$,
 $1$ at the state in $\mathcal{B}$, and $\bot$ at the states in $\mathcal{I}$ and other states. That is  $s_1(B)=0$,
 $s_2(B)=0$, $s_4(B)=1$, and $s_i(B)=\bot$ where $s_i\in S$ and $i\not=1$, $2$, or
 $4$. Subsequently, by making $V_V'=V_V\cup\{B\}$ and $V_I'=V_I$,
 the failure state is separated into three states in the refined abstract model  as illustrated in
 Figure \ref{fig:refine}.
 Note that, only the failure state is
 separated into three states, and other states are the same as in
 the abstract model. Especially, when $\mathcal{I}=\emptyset$, the failure
 state is separated into two new states.

Therefore, given a failure state $s_i$ (as well as $\mathcal{D}$,
$\mathcal{B}$ and $\mathcal{I}$) in the abstract model
$K=(S,S_0,R,L)$ where $S\subseteq \Sigma= \Sigma_{v_1}\times ...
\times\Sigma_{v_n}$ and $V_V=\{v_1,...,v_n\}$, to obtain the
abstract model $\hat{K}=(\hat{S},\hat{S_0},\hat{R},\hat{L})$, a
boolean variable $B$ is added as a visible variable with $s(B)=0$ if
$s\in \mathcal{D}$, $s(B)=1$ if $s\in \mathcal{B}$, and $s(B)=\bot$
if $s\not\in (\mathcal{D}\cup \mathcal{B})$. Thus, the set of all
possible states in the refined abstract model will be:
$$\hat{\Sigma}=\Sigma\times\Sigma_{B}$$
where $\Sigma_{B}=\{B=d\mid d\in \{0,1,\bot\}\}$. Accordingly, the
refined abstract model $\hat{K}=(\hat{S},\hat{S_0},\hat{R},\hat{L})$
can be obtained by Algorithm {\scshape Refine}.

\begin{algorithm}[h]
\caption{: {\scshape Refine}$(K, \mathcal{D}, \mathcal{B},
\mathcal{I}, B)$}
  {\bf Input}: the abstract model $K=(S,S_0,R,L)$ with $V_V$ being visible; $\mathcal{D}$, $\mathcal{B}$ and $\mathcal{I}$ reported by Algorithm {\scshape CheckSpurious}; the new boolean variable $B$ which will be added\\
  {\bf Output}: the refined model $\hat{K}=(\hat{S},\hat{S_0},\hat{R},\hat{L})$

    \begin{algorithmic}[1]
\STATE    $s(B)=0$ if $s\in \mathcal{B}$; $s(B)=1$ if $s\in
\mathcal{D}$; $s(B)=\bot$ if $s\not\in \mathcal{D}\cup\mathcal{B}$;
\STATE  $\hat{S}=\{\hat{s}\in \hat{\Sigma}\mid\mbox{ there exists
}s\in S\mbox{ such that }h(s,V_V\cup B)=\hat{s}\}$;

\STATE $\hat{S_0}=\{\hat{s}\in \hat{S}\mid\mbox{ there exists }s\in
S_0\mbox{ such that }h(s,V_V\cup B)= \hat{s}\}$;

\STATE $\hat{R}=\{(\hat{s_1},\hat{s_2})\mid \hat{s_1},\hat{s_2}\in
\hat{S}, \mbox{and there exist }s_1,s_2\in S\mbox{ such that
}h(s_1,V_V\cup B)=\hat{s_1}, h(s_2,V_V\cup B) = \hat{s_2} \mbox{ and
}(s_1,s_2)\in R\}$;

\STATE $L(\hat{s})=\bigcup\limits_{s\in S, h(s,V_V\cup B)=\hat{s}}
L(s)$;

\STATE return $\hat{K}=(\hat{S},\hat{S_0},\hat{R},\hat{L})$;
    \end{algorithmic}
\end{algorithm}


It can be observed that, the new refinement algorithm is linear to
the size of the state space, since it only needs to assign to the
new added boolean variable at each state. Further, in each
iteration, at most two more states are added (only one node is added
when $\mathcal{I}$ is empty). With the algorithm by choosing some
invisible variable visible, when $\mathcal{D}$ and $\mathcal{B}$ are
separated, other nodes (usually a huge number in the real systems in
practise) will also be separated. To illustrate the intrinsic
property of the new refining algorithm, a simple example is given
below.

\begin{Exm}\rm
The Kripke structure illustrated in l.h.s of Figure
\ref{fig:compare} (1) presents an original model where three
variables $x_1$, $x_2$ and $x_3$ are involved. Assume that $x_2$ and
$x_3$ are insensitive to the property which is expressed in a
temporal logic formula. Thus, by making $x_2$ and $x_3$ invisible,
the abstract model can be obtained by Algorithm {\scshape Abstract}
as illustrated in the r.h.s of Figure \ref{fig:compare} (1).
\begin{figure}[htp]
\centerline{\includegraphics[height=6.6cm]{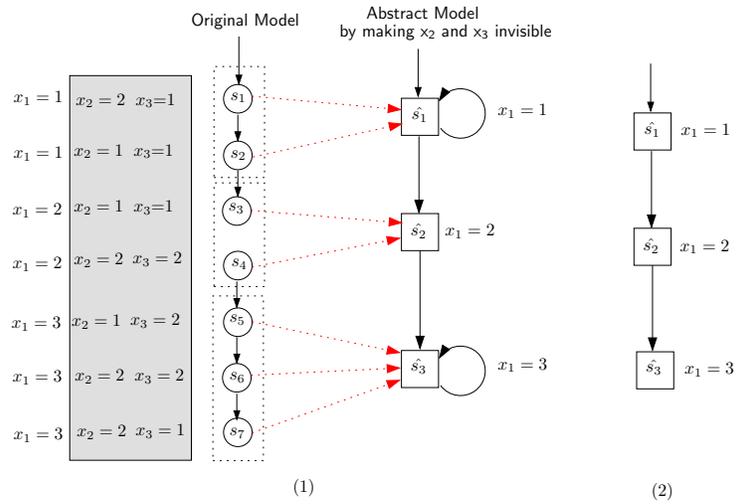}}\caption{Abstraction
by making $x_2$ and $x_3$ invisible}\label{fig:compare}
\end{figure}

Suppose that a counterexample is found by a model checker as
depicted in Figure \ref{fig:compare} (2). Then, by Algorithm
{\scshape CheckSpurious}, it will report that $\hat{s_2}$ is a
failure state, and $\mathcal{D}=\{s_3\}$, $\mathcal{B}=\{s_4\}$.
First, we show the refined abstract models by the method in the
related works \cite{Clarke04,CGHS02,HSGS09,HSNH10}. The refined
abstract model obtained by making $x_2$ and $x_3$ visible are
illustrated in Figure \ref{fig:compare1} (1) and (2) respectively.
\begin{figure}[htp]
\centerline{\includegraphics[width=12.5cm]{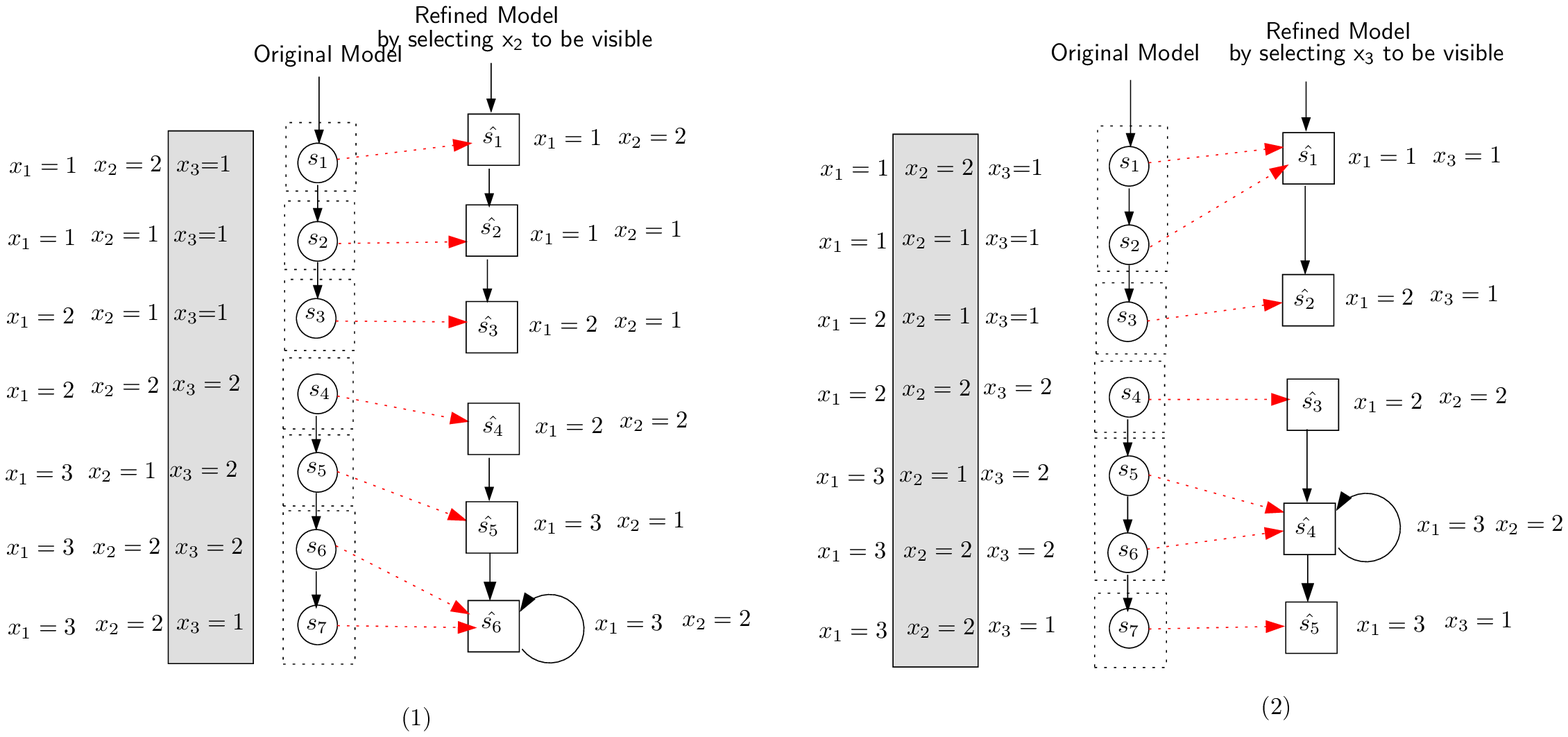}}\caption{Refinement
by the old algorithm}\label{fig:compare1}
\end{figure}
It can be observed that the one by making $x_3$ visible is the
smallest refined model under the method by making some invisible
variables visible. Clearly, to find the coarsest refined model, in
this way, is an NP-hard problem.

By our method, as depicted in Figure \ref{fig:compare2},
\begin{figure}[htp]
\centerline{\includegraphics[width=8.5cm]{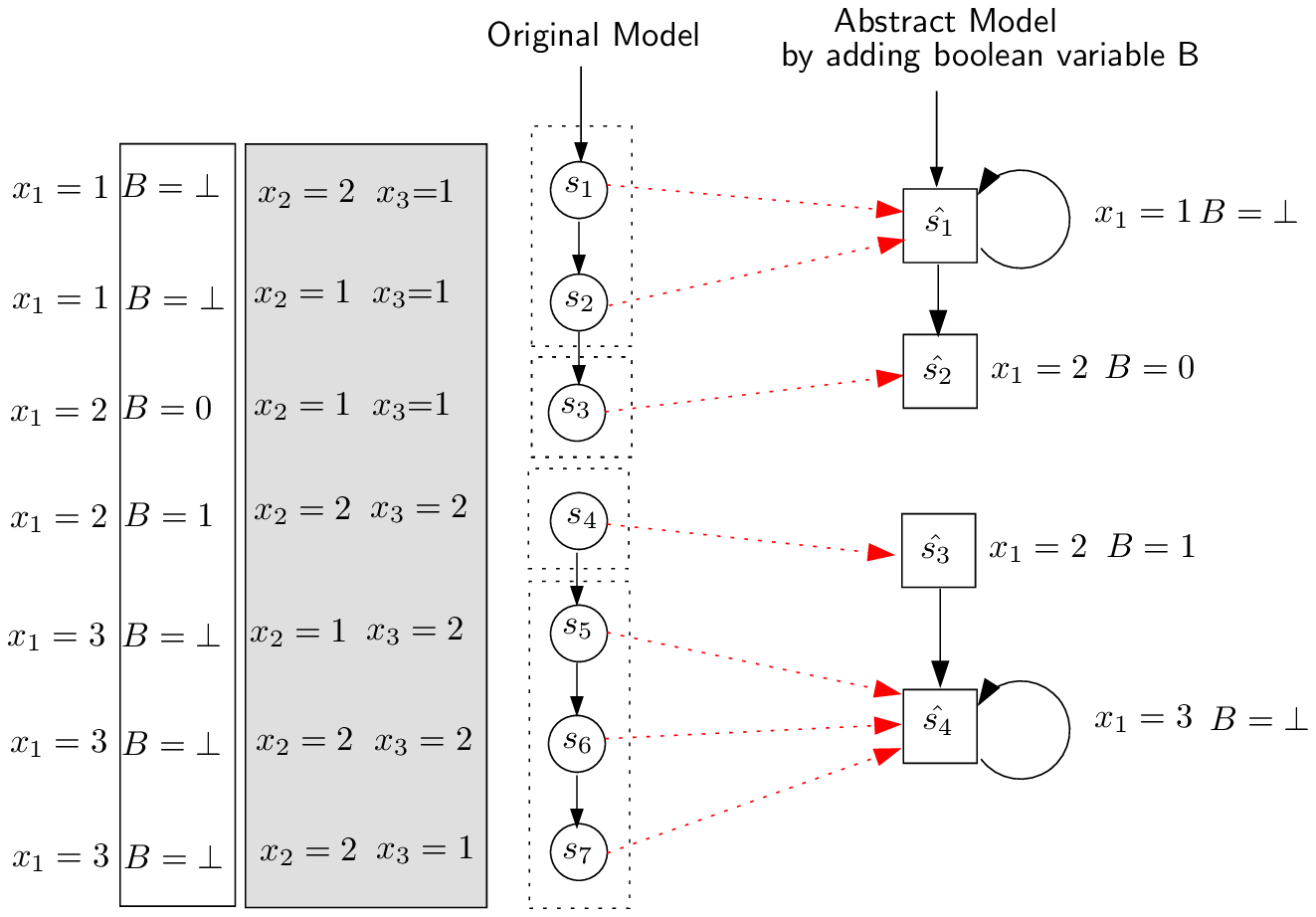}}\caption{Refinement
by the new algorithm}\label{fig:compare2}
\end{figure}
a new boolean variable $B$ is added to the system and made visible.
Then the refined abstract model is obtained where only the failure
state is separated into two states with other states unchanged.
Clearly, the new refining algorithm avoids the NP-hard problem for
finding the smallest set of visible variables. Moreover, the new
refined abstract model is smaller than the best result produced in
the method by further making some invisible variables visible.
\hfill{$\Box$}
\end{Exm}

\begin{figure}[htp]
\centerline{\includegraphics[height=3.6cm]{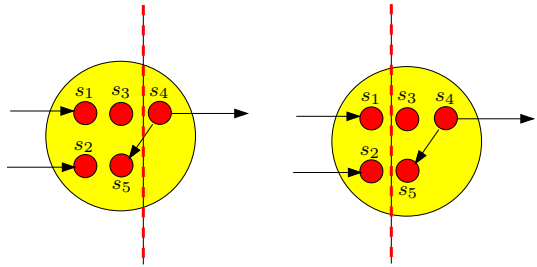}}\caption{Smallest
refinement}\label{fig:smallest1}
\end{figure}

Clearly, the refined model obtained by Algorithm {\scshape Refine}
is not the smallest one. And the smallest refined abstract model can
be easily obtained by assigning the new added variable $B$ by $0$ or
$1$ at the states in $\mathcal{I}$, i.e. the failure state is
separated into $\mathcal{D}\cup\mathcal{I}$ and $\mathcal{B}$, or
$\mathcal{D}$ and $\mathcal{B}\cup\mathcal{I}$. This is intuitively
presented in Figure \ref{fig:smallest1}. Compared to Algorithm
{\scshape Refine}, only one state is saved in the refinement.
However, more iterations will be introduced into the abstract model
checking since $\mathcal{D}\cup\mathcal{I}$ or
$\mathcal{B}\cup\mathcal{I}$ may be found as a failure state
further.

\section{Abstract Model Checking Framework }
With the new proposed algorithms, the abstract model checking
framework is presented. First, the abstract model is obtained by
Algorithm {\scshape Abstract}. Then a model checker is employed to
check whether or not the abstract model satisfies the desired
property. If no errors are found, the model is correct. However, if
a counterexample is reported, it is checked by Algorithms {\scshape
CheckSpurious}.
\begin{algorithm}[h]
\caption{: {\scshape AbstractMC}}
  {\bf Input}: A model $K=(S,S_0,R,L)$ in Kripke structure, and a desired property $\phi$ in temporal logic\\
  {\bf Output}: a counterexample that violates $\phi$

    \begin{algorithmic}[1]
\STATE {\bf Initialization}: $int$ $i=1$;
        \STATE $\hat{K}=${\scshape Abstract}$(K, V_I)$;
        \STATE $MC(\hat{K}, \phi)$;
        \WHILE {a counterexample $\hat{\Pi}$ is found}
            \STATE  {\scshape CheckSpurious}$(\hat{\Pi})$;
            \STATE  {\bf if }$\hat{\Pi}$ is a real counterexample, return $\hat{\Pi}$; break;
            \STATE  {\bf else} $\hat{K}=${\scshape Refine}$(\hat{K}, \mathcal{D},\mathcal{B}, \mathcal{I}, B_i)$; $i=i+1$; $MC(\hat{K}, \phi)$;
        \ENDWHILE \label{code:recentEnd}
        \STATE {\bf if} no counterexample is found, $K$ satisfies $\phi$.
    \end{algorithmic}
\end{algorithm}
If the counterexample is not spurious, it will be a real
counterexample that violates the system; otherwise, the
counterexample is spurious, and  Algorithm {\scshape Refine} is used
to refine the abstract model by adding a new visible boolean
variable $B$ to the system. Then the refined abstract model is
checked with the model checker again until either a real
counterexample is found or the model is checked to be correct. This
process is formally described in Algorithm {\scshape AbstractMC}
where a subscript $i$ is used to identify different boolean
variables that are added to the system in each refinement process.
Initially, $i$ is assigned by $1$. After each iteration of Algorithm
{\scshape Refine}, $i$ is increased by $1$. Basically, finitely many
boolean variables will be added since the systems to be verified
with model checking are finite systems.

\section{Conclusion}
 An efficient method for abstraction refinement is given in this
 paper. With this approach, the NP-hard state separation problem can be avoided, and the
 smaller refined abstract model can also be obtained. This can improve the abstract based model checking,
  especially the counterexample guided abstraction refinement model checking. In the near future,
  the proposed  algorithm will be implemented and integrated into the tool
  CEGAR. Further, some case studies will be conducted to evaluate
  the algorithm.


\begin{thebibliography}{99}


\bibitem{Clarke81} \textsc{E.M.Clarke and E.A.Emerson}. \emph{Desigh and syntesis of of synchronization skeletons using
branching time temporal logic}. In Logic of Programs: Workshop, Yorktown Heights, NY,
May 1981, LNCS 131, Springer, 1981.

\bibitem{QS82}\textsc{J.P.Quielle and J.Sifakis}. \emph{Specification and verification of concurrent systems in CESAR}. In
Proceedings of the 5th international symposium on programming, pp.337-350, 1981.

\bibitem{CGJLV00} \textsc{E.Clarke, O.Grumberg, S.Jha, Y.Lu, and H.Veith}, \emph{Counterexample guided
abstraction refinement}, in Proc. 12th Int. Conf. Computer-Aided
Verification (CAV¡¯00), vol. 1855, E. Emerson and A. Sistla, Eds. New
York, 2000.

\bibitem{Krip63}
\textsc{S.A.Kripke}. \emph{Semantical analysis of modal logic I: normal
propositional calculi}, Z. Math. Logik Grund. Math. 9, 67-96, 1963.

\bibitem{Clarke04} \textsc{E.M.Clarke}. \emph{Sat Based Counterexample-Guided Abstraction-Refinement}. IEEE Trans. Computer Aided Design,
vol.23, no. 7, pp. 1113-1123, July 2004.

\bibitem{HSHGS10}\textsc{Fei He, Xiaoyu Song, William N. N. Hung, Ming Gu, Jiaguang Sun}. \emph{Integrating Evolutionary Computation with Abstraction Refinement for
Model Checking}. IEEE Trans. Computers 59(1): 116-126 (2010)


\bibitem{Rushby99}\textsc{J. Rushby}. \emph{Integrated formal verification: Using model checking with
automated abstraction, invariant generation, and theorem proving}.
presented at Theoretical and Practical Aspects of SPIN Model
Checking: Proc. 5th and 6th Int. SPIN Workshops. [Online].
Available: citeseer.nj.nec.com/rushby99integrated.html


\bibitem{HSGS09}\textsc{Fei He, Xiaoyu Song, Ming Gu, Jia-Guang Sun}. \emph{Heuristic-Guided Abstraction Refinement}. Comput. J. 52(3): 280-287 (2009)

\bibitem{HSNH10}\textsc{Fei He, Xiaoyu Song, William N. N. Hung, Ming Gu, Jiaguang Sun}. \emph{Integrating Evolutionary Computation with Abstraction Refinement for Model Checking}. IEEE Trans. Computers 59(1): 116-126 (2010)

\bibitem{Krushan94}\textsc{R.P.Kurshan}. \emph{Computer Aided Verificaton of Coordinating
Processes.} Princeton Univ. Press, 1994.

\bibitem{WLJHS06} \textsc{C. Wang, B. Li, H. Jin, G.D. Hachtel, F. Somenzi}. \emph{Improving
Ariadne's Bundle by Following Multiple Threads in Abstraction
Refinement}. IEEE Trans. Computer Aided Design, vol. 25, no. 11, pp.
2297-2316, Nov. 2006.


\bibitem{CGHS02} \textsc{E.M. Clarke, A. Gupta, J.H. Kukula, and O. Strichman}. \emph{SAT Based
Abstraction-Refinement Using ILP and Machine Learning Techniques}.
Proc. Computer-Aided Verification (CAV), E. Brinksma and K.G.
Larsen, eds., pp. 265-279, 2002.

\bibitem{CCKSVW02} \textsc{P. Chauhan, E.M. Clarke, J. Kukula, S. Sapra, H. Veith, and D.
Wang}. \emph{Automated Abstraction Refinement for Model Checking
Large State Spaces Using SAT Based Conflict Analysis}. Proc. Formal
Methods in Computer-Aided Design (FMCAD), 2002.

\bibitem{HJMS02} \textsc{T.A. Henzinger, R. Jhala, R. Majumdar, and G. Sutre}. \emph{Lazy
Abstraction}. Proc. Symp. Principles of Programming Languages, pp.
58-70, 2002.

\bibitem{GKMFV03} \textsc{M. Glusman, G. Kamhi, S. Mador-Haim, R. Fraer, M.Y.
Vardi}. \emph{Multiple-Counterexample Guided Iterative Abstraction
Refinement: An Industrial Evaluation}. Proc. Tools and Algorithms
for the Construction and Analysis of Systems (TACAS), pp. 176-191,
2003.

\bibitem{GD00} \textsc{S.G. Govindaraju, D.L. Dill}. \emph{Counterexample-Guided Choice of
Projections in Approximate Symbolic Model Checking}. Proc. Int¡¯l
Conf. Computer-Aided Design (ICCAD), pp. 115-119, 2000.
\end{thebibliography}
\end{document}